\documentstyle[prd,aps,floats]{revtex}

\def\be{\begin{equation}}
\def\ee{\end{equation}}
\def\ben{\begin{eqnarray}}
\def\een{\end{eqnarray}}

\begin{document}

\input epsf
\renewcommand{\topfraction}{1.0}
\twocolumn[\hsize\textwidth\columnwidth\hsize\csname
@twocolumnfalse\endcsname

\title{Boson stars with generic self-interactions}

\author{Franz E.~Schunck${^1}$ and Diego F.~Torres${^{2}}$}

\address{${^1}$Institut f\"ur Theoretische Physik, Universit\"at zu
K\"oln, 50923 K\"oln, Germany}
\address{${^2}$Departamento de F\'{\i}sica, Universidad Nacional
de La Plata, C.C. 67, 1900 La Plata, Buenos Aires, Argentina}

\date{\today}

\maketitle

\begin{abstract}
We study boson star configurations with generic, but not
non-topological, self-interaction terms, i.e.~we do not restrict
ourselves just to consider the standard $\lambda |\psi|^4$
interaction but more general U(1)-symmetry-preserving profiles. We
find that when compared with the usual potential, similar results
for masses and number of particles appear. However, changes are of
order of few percent of the star masses. We explore the stability
properties of the configurations, that we analyze using
catastrophe theory. We also study possible observational outputs:
gravitational redshifts, rotation curves of accreted particles,
and lensing phenomena, and compare with the usual case.
\end{abstract}

\vskip2pc]
\newpage
\section{Introduction}

In the past two decades, the study of the universe in its earlier
stages has given an important role to scalar fields: conventional
models for inflation rely on its possible potential energy and
topological defects may form when a scalar field breaks some
fundamental symmetry. Both of these processes may be responsible
for the formation of the large scale structure we now see. In
addition, scalar fields connected with time or space-time
variation of fundamental constants have produced the most powerful
alternative theories of gravitation, known as scalar-tensor
theories, and entered in all unification scheme trials made so
far. It must be said, however, that no fundamental cosmological
scalar field has been directly observed yet.

If one accepts the need of scalar fields in the cosmological
scenario, one interesting question naturally arise: may these
scalars be the seed of astrophysical structures or of observable
phenomena that could signal their existence? Objects made up of
scalar massive particles were introduced as early as 1968 by Kaup
\cite{KAUP} and Ruffini and Bonazzola \cite{RB}. These
configurations are now known as boson stars. These stars, contrary
to the more common neutron or fermion stars, are not supported by
Pauli's exclusion but by Heisenberg's uncertainty principle,
which effectively keeps scalars from being localized to within
their Compton wavelength and prevents their collapse to a black
hole.

The first investigations with a nonlinear $\psi^6$ potential
were carried out by Mielke and Scherzer \cite{MS81}. They calculated this
potential from a nonlinear Heisenberg-Pauli-Weyl spinor equation
and found the first scalar field solutions with nodes.
More recently, Colpi et al.~\cite{COLPI}
proved that the existence of a self-interaction in the
boson Lagrangian could yield higher values for the masses of the
configurations.
This means that for some values of the free parameters existing
in the theory, stellar
structures of appreciable masses and extremely high density may arise. This
could have incredible effects on our understanding of
the non-baryonic content of the universe
and have different observational effects:
affecting usual objects as galaxies \cite{SCHUNCK}
or even light trajectories in powerful, degree-ranged, microlensing
phenomena \cite{DABROWSKI}.
These kind of features, as well as the important
point of star rotation (see \cite{RBS} among others),
was thoroughly studied in the past
few years, including gauge charged stars,
boson-fermion models (see \cite{reviews} for reviews)
and models of stars in alternative
theories of gravity \cite{TORRES-BOSON}.

In all these works, the self-interaction was always choiced to be
like $\lambda |\psi|^4$, thus giving a matter sector provided by,
\begin{equation}
{\cal L}_{{\rm m}} = -\frac{1}{2} g^{\mu \nu} \, \partial_\mu \psi^*
        \partial_\nu \psi -\frac{1}{2} m^2 |\psi|^2
        -\frac{1}{4} \lambda |\psi|^4 \,,
\end{equation}
where $\lambda$ is a constant. However, a priori, there is no
reason to maintain just the form adopted for the self-interaction
and not a more generic potential term $V(\psi)$. Although in order
to retain the $U(1)$ symmetry of the whole Lagrangian, this
potential must be a function of $|\psi|^2$, in principle, it may
have any other particular form. In this work we would like to
explore whether substantial variations in the form of the
potential may yield appreciable changes in the final
configurations, the binding energies, or the masses of boson
stars. We also analyze the changes that a different potential
introduces in gravitational redshifts and microlensing phenomena.
This could be important to discover if different scenarios for
dark matter candidates, explanation of galactic rotation curves
\cite{SCHUNCK}, or new observational gravitational effects may
appear \cite{LIDDLE-SCHUNCK}. Additionally, new potentials may
strongly affect gravitational memory and evolution scenarios
\cite{MEMORY} within theories with time variation of Newton's
constant (for the most recent account of this see
Ref.~\cite{WHINNETT}).

Knowledge about that modifications in the boson potential may
yield to profound changes in the structure of the stellar objects
comes from the works of Friedberg, Lee, and Pang
\cite{L87,FLP87,lee} on non-topological soliton stars. However, we
point out that a systematic study on their possible observational
signatures is still absent. We shall briefly mention their
features and compare with our potential choices below. However,
{\it our main aim here is to study if potentials which are not
non-topological ones may still introduce significant changes in
the configurations.}

This work is organized as follows. In the next section we
introduce the models for boson stars and search, with different
self-interaction terms, for their configurations and stability
properties. Section III compares the results of these generic
boson star models, concerning observational outputs, with those
obtained for the usual $\lambda |\psi|^4$ term.
We give our conclusions and provide a brief discussion in the
final section.

\section{The models}

We shall take into account the following boson Lagrangian
\begin{equation}
\label{gen}
{\cal L}_{{\rm m}} = -\frac{1}{2} g^{\mu \nu} \, \partial_\mu \psi^*
        \partial_\nu \psi -\frac{1}{2} U(|\psi|^2) \, ,
\end{equation}
with the potential
\begin{equation}
\label{pot}
U(|\psi|^2) = m^2 |\psi|^2 + \frac{1}{2} V(|\psi|^2) \,.
\end{equation}
This action possesses an invariance under the global $U(1)$ transformation
$\psi \rightarrow e^{i\theta}\psi$, which gives rise to a conserved current
\begin{equation}
J^\mu=ig^{\mu\nu} \left( \psi^* \partial_\nu \psi - \psi \partial_\nu
\psi^* \right),
\end{equation}
and a corresponding conserved charge
\begin{equation}
N=\int dx^3 \sqrt{-g} J^0.
\end{equation}
This conserved quantity can be identified with the number of bosons present
in the stellar structure.
We shall consider the usual spherically symmetric metric
\begin{equation}
\label{metric}
ds^2=-B(r) dt^2 + A(r) dr^2 +r^2 d\Omega^2,
\end{equation}
and shall also demand a spherically symmetric
form for the field which describe the boson, {\it i.e.} we adopt the ansatz:
\begin{equation}
\label{boson}
\psi(r,t)=\chi(r) \exp{[-i\varpi t]}.
\end{equation}
This ensures that a static solution is still possible.
To establish that this time dependence is the lowest energy solution at
a fixed number of particles, it is necessary to make a first order
variation $\delta (E-\varpi N)$, where in general $\varpi$ is a
Lagrangian multiplier associated with the conservation of $N$ and $E$
is the energy (mass) of the system. See for instance the appendix
of Ref.~\cite{FLP87} for a detailed derivation. This form for
$\psi$ is independent of the potential $U$.
For a more detailed derivation of the basic model we refer the reader to
Refs.~\cite{KAUP,RB,COLPI,reviews}.
Using the Einstein gravitational action we obtain the field equations,
\begin{eqnarray}
\label{f1}
\frac{2M^\prime}{x^2} & = & \sigma^2 \left( \frac{\Omega^2}{B} +1 \right)
 +  \frac{\sigma^{\prime 2}}{A} + \nonumber \\
& & \quad
\frac12 \left(
\frac {4 \pi}{m^2 M_{{\rm Pl}}^2} \right) V(\sigma M_{{\rm Pl}} /
\sqrt{4 \pi}),
\end{eqnarray}

\begin{eqnarray}
\label{f2}
\frac{B^\prime}{ABx} & - & \frac{1}{x^2} \left( 1-\frac 1A \right)  = \sigma^2
\left( \frac{\Omega^2}{B} -1 \right) +  \nonumber \\
& &  \quad \frac{\sigma^{\prime 2}}{A} - \frac 12 \left(
\frac {4 \pi}{m^2 M_{{\rm Pl}}^2} \right) V(\sigma M_{{\rm Pl}} /
\sqrt{4 \pi}),
\end{eqnarray}
and the Klein-Gordon equation for $\psi$,

\begin{eqnarray}
\label{f3}
\sigma^{\prime\prime} & + & \sigma^\prime
\left( \frac 2x - \frac{A^\prime}{2A} +
\frac{B^\prime}{2B} \right) +
A \sigma
\left( \frac{\Omega^2}{B} -1 \right) \nonumber \\
&&\quad
- \frac 14
\left( \frac {4 \pi}{m^2 M_{{\rm Pl}}^2} \right)
\frac{dV(\sigma M_{{\rm Pl}} / \sqrt{4 \pi})}{d\sigma}=0.
\end{eqnarray}
In these equations, we have used dimensionless units, which are
common to those introduced in the paper by Colpi et al.,
\begin{equation}
\label{x}
x=mr \,,
\end{equation}
for the radial coordinate ($\,{}^\prime$ stands for the derivatives
with respect to $x$) and,
\begin{equation}
\label{dimensionless}
\Omega=\frac{\varpi}{m},\;\;
\sigma=\sqrt{4\pi} \frac{\chi(r)}{M_{{\rm Pl}}},
\end{equation}
where $M_{{\rm Pl}} \equiv G^{-1/2}$ is the Planck mass.

In order to consider the total amount of mass contained within
a radius $x$ we have changed the function $A$ in the metric to
its Schwarzschild form,
\begin{equation}
\label{M}
A(x)=\left(1-\frac{2M(x)}{x}\right)^{-1}.
\end{equation}
Note that the terms corresponding to the potential are correctly
normalized. From the Lagrangian (\ref{gen}) one may see that $V$
has dimensions of [Energy]$^4$, and all of them are divided by two
squares of masses. However, in order to avoid the explicit
appearance of the boson mass $m$ and the Planck mass $M_{{\rm
Pl}}$, we need to define the form of $V$. We shall look at several
choices.\\

The first expansion to nonlinear potentials can be found in the 1981
paper by Mielke \& Scherzer \cite{MS81}. They constructed a potential
for the Klein-Gordon equation
from the Heisenberg-Pauli-Weyl nonlinear spinor equation. It has the general
form $U=m^2 |\psi|^2 - \alpha_1 |\psi|^4 + \alpha_2 |\psi|^6$, where
$\alpha_1$ and $\alpha_2$ are two positive constants. They presented
solutions with nodes for the first time.

The standard (CSW \cite{COLPI}) choice is
$V(\psi)=\lambda |\psi|^4$, with $\lambda$ a constant.
Here, the usual adimensionalization appears,
\begin{eqnarray}
\frac{4\pi}{m^2 M_{{\rm Pl}}^2}
V(\sigma M_{{\rm Pl}} / \sqrt{4 \pi}) =\;\;\;\;\;\;  \mbox{} \nonumber \\
\frac{4\pi}{m^2 M_{{\rm Pl}}^2}
\sigma^4 \lambda \frac{M_{{\rm Pl}}^4}{(4\pi)^2}=
\lambda \frac{ M_{{\rm Pl}}^2}{4\pi m^2} \sigma^4= \Lambda\sigma^4,
\end{eqnarray}
where $\Lambda=\lambda  M_{{\rm Pl}}^2/ 4\pi m^2 $.
As we stated in the introduction, with this choice, the order of magnitude
of boson star masses is deeply enhanced.
It grows from $M \sim M_{{\rm Pl}}^2/m$ when $\Lambda=0$ to
$M \sim M_{{\rm Pl}}^3/m^2$ when
$\Lambda \neq 0$. Recall that the mass of a neutron star is roughly given
by the Chandrasekhar mass $M_{Ch} \sim M_{{\rm Pl}}^3/m_n^2$ which is close to
a solar mass ($m_n$ is the neutron mass).

Other three options we would like to explore are (note that these are
options for $U(|\psi|^2)$, given in Eq.~(\ref{pot}), not only for
$V(|\psi|^2)$ as will be clear below):

\begin{itemize}

\item Cosh-Gordon potential:\\
$$
U_{\cosh }=
\alpha m^2 \bigl[\cosh(\beta \sqrt{|\psi|^2}) - 1\bigr]
\nonumber
$$

\item Sine-Gordon potential:
\begin{eqnarray*}
U_{\sin } &=&
\alpha  m^2 \bigl[\sin(\pi/2 [\beta \sqrt{|\psi|^2}-1])+1\bigr]  \\
&=& \alpha m^2 \bigl[ 1 - \cos (\pi/2 \beta \sqrt{|\psi|^2} ) \bigr]
\end{eqnarray*}

\item $U(1)$-Liouville potential:
$$
U_{\exp }=
\alpha m^2 \bigl[\exp(\beta^2 |\psi|^2)-1\bigr]
$$
The usual Liouville potential $\exp(\beta \psi )$ has to be changed so that
a $U(1)$ symmetry is ensured.

\end{itemize}
Let us first consider a series expansion of these potentials. In
order to do so we shall consider a value of $\beta$ such that,
when going from $\psi$ to the dimensionless $\sigma$, the
arguments of the functions are not affected. The parameter $\beta$
is arbitrary, it enlarges the parameter space of the solutions, as
was the case with $\Lambda$ in CSW's solutions. The appearence of
the factor $\beta$ is just because dimensional grounds, while
$\alpha$ can be used to get a simple mass term in the boson
Lagrangian, as we shall see below. Taking this into account, the
series expansions are,

\begin{eqnarray}
U_{\cosh } &=& \alpha m^2
\Bigl[ \cosh (\beta \sigma) - 1 \Bigr] =
\alpha m^2 \times \nonumber \\
&& \left[ \frac{\beta^2 \sigma^2}{2}
+\frac{\beta^4 \sigma^4}{24}+
\frac{\beta^6 \sigma^6}{720}+
\frac{\beta^8 \sigma^8}{40320} + \ldots \right] \; ,
\end{eqnarray}

\begin{eqnarray}
&& U_{\sin} = \alpha m^2 \Bigl[ \sin (\pi/2 \; (\beta \sigma - 1
)) + 1  \Bigr] = \alpha m^2 \times \nonumber \\ && \left[
\frac{\beta^2 \pi^2 \sigma^2}{8} -\frac{\pi^4 \beta^4
\sigma^4}{384}+ \frac{\pi^6 \beta^6 \sigma^6}{46080}- \frac{\pi^8
\beta^8 \sigma^8}{10321920} + \ldots \right] \; ,
\end{eqnarray}

\begin{eqnarray}
U_{\exp} &=& \alpha m^2
\Bigl[ \exp (\beta^2 \sigma^2) - 1 \Bigr ] =
\alpha m^2 \times \nonumber \\
&&\left[ \beta^2 \sigma^2 + \frac{1}{2} \beta^4 \sigma^4 + \frac{1}{6} \beta^6
  \sigma^6 + \frac{1}{24} \beta^8 \sigma^8 + \ldots  \right] \; .
\end{eqnarray}

Note that, from each expansion, we are recognizing a usual mass
term (proportional just to $m^2$). This term is very important:
without it, it is impossible to find solutions with exponential
decrease of the scalar field, something relevant for the
definition of the star radius. Then, in order to be consistent
with equations (\ref{pot}) and (\ref{f1}-\ref{f3}), the field
equations and the definition of $U$, and to avoid a useless double
counting of the mass term, we must take particular choices for
$\alpha$; the parameter $\beta$ is still free for choice.
\begin{itemize}

\item Cosh-Gordon potential:
$$
\alpha = 2 (M_{{\rm Pl}}^2/4\pi)/\beta^2
$$

\item Sine-Gordon potential:
$$
\alpha = (8/\pi^2) (M_{{\rm Pl}}^2/4\pi)/\beta^2
$$

\item $U(1)$-Liouville potential:
$$
\alpha = (M_{{\rm Pl}}^2/4\pi)/\beta^2
$$

\end{itemize}

In this way, the potential $V$ is everything but the first terms
in each of the previous series. Hence, it is a series of
attractive-repulsive self-interactions in the case of the
Sine-Gordon potential and a series of repulsive potentials in the
Cosh-Gordon case. The case of the $U(1)$-Liouville potential is
reminiscent of the Cosh-Gordon one, in the sense of being a series
of repulsive power law self-interactions, just the coefficients
differ. The form of these potentials and others mentioned above
are shown in Fig.~1.

For the numerical procedure, it is best to make a redefinition of
the scalar field mass
\be
\widetilde m^2 = \alpha m^2 \; , \ee and with this also redefine
the coordinate $x$ and the frequency $\Omega$. Notice that $\beta$
still appears within the differential equations while $\alpha$
does not.

\subsection{Soliton stars}

A potential with symmetry breaking was investigated by Lee et
al.~\cite{L87,FLP87,lee}. They called the solutions non-topological
soliton stars, and found
that the mass has units of $M_{\rm Pl}^4/(m \sigma_0^2)$ which is huge in
comparison with a boson or neutron star (for the case of comparable
boson and fermion masses). The potential investigated was
$U=m^2 |\psi|^2 (1 - |\psi|^2/\sigma_0^2)^2$ where
$\sigma_0$ is a constant; this belongs to the more general forms of potentials
derived in \cite{MS81}.

Compared with the usual boson star case, non-topological
soliton stars have to fulfill two characteristics:
\begin{enumerate}
\item  The Lagrangian must be invariant under a global $U(1)$ transformation.
\item  In the absence of gravity, the theory must have non-topological
  solutions; i.e.~solutions with a finite mass, confined to a finite region
  of space, and non-dispersive.
\end{enumerate}
In general, boson stars accomplish the requirement 1.~but not 2.
Invariance under $U(1)$ only requires that the potential be a
function of $\psi^* \psi$, and in order to fulfill condition 2.,
$U$ must contain attractive terms. This is why the coefficient of
$(\psi^* \psi)^2$ of Lee's potential has a negative sign. Finally,
when $|\psi| \rightarrow \infty$, $U$ must be positive, which
leads, minimally, to a sixth order function of $\psi$ for the
self-interaction. It is then clear that CSW's, $U_{{\cosh }}$, and
$U_{{\rm exp}}$ choices are not non-topological potentials.
Neither of them have attractive terms. This is why, a priori, we
may say that the order of magnitudes for the boson star masses
remains the same as in CSW's case. The Sine-Gordon potential
$U_{{\rm sin}}$ has, on the contrary, a similar series expansion,
up to the sixth order, to that corresponding to Lee's potential.
But here, what happens is that $U_{{\rm sin}}$, in the absence of
gravity and for a real scalar field, has not a non-topological soliton
solution, instead, it
has a topological one. It has a degenerate vacuum: an infinite set
of $\sigma$ values for which $U_{{\rm sin}}=0$. For a detailed
account of this, we refer the reader to Lee's book, especially
Chapter 7 and exercise 7.1 \cite{BOOK}. Then, $U_{{\rm
sin}}$ neither is a non-topological soliton potential.

\subsection{Numerical solutions}

Fig.~2 shows the usual plot of boson star configurations for the case
of the Cosh-Gordon potential. The maximal mass is slightly higher than in the
standard case due to the additional higher-order repulsive terms in the
potential.
For $\beta =1$, we find $M_{\rm max}=0.638$ $M_{\rm Pl}^2/m$ and
$N_{\rm max}=0.658$ $M_{\rm Pl}^2/m^2$ what for $M$ and $N$ is higher
by 0.5\%.
Fig.~3 shows the stability analysis,
which can be done using catastrophe theory
\cite{KUSMARTSEV,KUSMARTSEV2,KUSMARTSEV3}. One necessary condition for the
configurations to be stable is a negative binding energy. However, this is
not sufficient. Fig.~3 shows the appearence of two cusps signaling that there
is a change in the star stability. The first branch is the only stable one,
while the second and third are both unstable.

Fig.~4 represents similar profiles, but for the Sine-Gordon
potential. In this case, the maximal values of mass and particle
number are below the pure mass potential case. The influence of
the higher order attractive terms is noticeably. For $M$ and $N$,
the maximal values are lower by about 2\%: $M_{\rm max}=0.620$
$M_{\rm Pl}^2/m$ and $N_{\rm max}=0.639$ $M_{\rm Pl}^2/m^2$.
Fig.~5 shows the bifurcation plot for this case. The diagram in
Fig.~5 shows cusps again, where is a change in stability. A remark
on the calculation of mass is in order. We check that the
calculation of the mass is correct by applying two different mass
definitions. The first is the Schwarzschild mass, which is defined
by the energy density $\rho $ and which also appears in the
asymptotic spherically symmetric space-time, $B(r)\rightarrow
1-2M/r$, where $M$ is the mass of the boson star. The formula for
the Schwarzschild mass is \ben M_S & = & 4\pi \int\limits_0^\infty
\rho r^2dr  \\
 & = & 4\pi \int\limits_0^\infty
    \left ( \varpi^2 \frac{\chi(r)^2}{B}+
    \left(\frac{d\chi(r)}{dr}\right)^2 \frac{1}{A}
       +U \right ) \; r^2\, dr  \, .
\een  A second mass formula can be derived for a general
quasi-static isolated mass insula where a time-like Killing vector
$\xi^\alpha =2n^\alpha $ exists \cite{RBS}. Tolman's expression
\cite{TOLMAN} is: \ben M_T = \int (2T_0^{\; 0}-T_\mu^{\; \mu })
         \sqrt{\mid g\mid} \; d^3x .
\een For an asymptotically flat spherically symmetric space-time,
both masses agree with each other.

Finally, Figs.~6 and 7 shows the behavior of the $U(1)$-Liouville
potential. It is both, qualitatively and quantitatively similar to
the usual CSW's case. We have stable and unstable branches. The
maximal mass and particle number are higher by about 5\% which are
the largest deviations from the standard case. The repulsive
potential terms yield larger contributions in comparison with the
Cosh-Gordon potential.

\section{Observational outputs}

At this stage, we have succeeded in proving that a change in the form of the
self-interaction among the bosons
yields appreciable -but small- changes in the form of the star configurations.
We shall discuss below the feasibility of detecting observational
consequences of these results concerning gravitational phenomena.
In particular, we are interested in seeing if appreciable differences
appear in the computation of gravitational redshifts, rotation curves, and
gravitational lensing features. For the Sine-Gordon potential, and because of
the smaller masses it produces, gravitational phenomena are diminished.
So we shall be mainly interested in Cosh-Gordon and $U(1)$-Liouville cases.

\subsection{Gravitational redshifts}

In this section, we follow Ref.~\cite{LIDDLE-SCHUNCK},
and make use of the assumption
that the scalar particles have no interaction
-other than gravitational- with baryonic matter.
Thus, this {\it normal} matter can penetrate the boson star
up to the center and if it emits radiation there,
well within the gravitational potential,
we expect the spectral features to be redshifted.

The gravitational redshift $z$ of our static line element is given by

\begin{equation}
1+z=\sqrt{ \frac{B(\infty)}{B(int)} },
\end{equation}
where $int$ stands for the position of the emitter particle with
respect to the star center. As the receiver is practically at
infinity, $B(\infty)\sim 1$. The maximum possible redshift is
obtained when the particle emits exactly at the center of the
boson star, where the metric deviates maximal from outside vacuum
space-time.

We are only interested in stable configurations, the maximum
redshift is then provided by the maximum value of $\sigma(0)$,
which gives the biggest mass. As it was shown in
Ref.~\cite{LIDDLE-SCHUNCK}, the simple mass term produced a
maximum redshift of 0.45 while for CSW's choice, with $\Lambda$
tending to infinity, one gets 0.69. Here we find $z_{max}=0.46$
for the Cosh-Gordon potential and $z_{max}=0.49$ for the
$U(1)$-Liouville potential. For comparison, we quote the result
$z_{max}=0.47$ for neutron stars.

\subsection{Rotation curves}

Another gravitational effect considered for CSW's choice
\cite{LIDDLE-SCHUNCK}, and which we would like to compare with the more
generic potentials here studied are the rotation curves of test particles
moving around boson stars. For our metric, circular
geodesics have a rotation velocity (as measured by an observer at
infinity) given by,
\begin{equation}
v_\varphi^2 = \frac{x B^\prime}{2}.
\end{equation}
In Fig.~8 we compare the rotation curves for the case $\Lambda=0$,
$\Lambda=100$ of CSW's choice, and our potentials: Cosh-Gordon and
$U(1)$-Liouville. Already from the usual case, it was shown that
the possible velocities that particles can reach are a notable
amount of the speed of light, and matter can have an impressive
kinetic energy. If such kinetic energy were transferred to
radiation, we could expect very luminous boson stars, orders of
magnitude more luminous than the Sun. This provides speculative
alternatives to accretion disks around black holes, and would make
boson stars almost indistinguishable from its final effects. This
is currently being analyzed, especially concerning the possibility
of having boson stars at the center of some galaxies, as was
proposed with neutrino balls. For the new potentials we are
analyzing here, we obtain that $U_{{\cosh }}$ and $U_{{\rm exp}}$
produce similar angular velocities to the $\Lambda=0$ case. Their
maximum velocity happens for values $x \sim 5$ and have typical
magnitudes of 100 000 km s$^{-1}$.

\subsection{Gravitational lensing}

Boson star microlensing effects were first investigated by
D\c{a}browski and Schunck \cite{DABROWSKI}, for CSW's $\Lambda=0$
case, also known as the mini-boson star. Their procedure, which we
closely follow, consist in studying the photon trajectories along
the curved (boson star generated) space-time. For particular
details of the derivation of quoted formulae see
Refs.~\cite{DABROWSKI,VIBRA}. In Ref.~\cite{VIBRA}, a related
study on microlensing features was made, taking a
scalar-field-generated naked singularity as lens. It has the
property of producing both, positive and negative binding angles;
in this later case, in a way similar to the recently studied
wormhole microlensing scenario \cite{TORR}.

The light traveling from a distant source is deflected, because of
the presence of the boson star, with a deflection angle given by
(see Fig.~\ref{fig9a} for a schematic drawing of the geometry),
\begin{equation}
\hat \alpha=\Delta \varphi - \pi,
\end{equation}
where,
\begin{equation}
\Delta \varphi = 2 \int_{r_0}^\infty \frac{dr}{r}
\frac{\sqrt{AB}}{\sqrt{r^2/b^2 -B}}, \label{deltaphi}
\end{equation}
with impact parameter $b=r_0 \sqrt{1/B(r_0)}$ ($r_0$ is the
closest distance between the light ray and the center of the boson
star: the first point where the square root in the denominator is
non-negative). The lens equation can be expressed as the
difference between the true angular position, $\beta$, and the
image positions, $\vartheta$, as \cite{SEF92,MICRO}
\begin{equation}
\sin(\vartheta - \beta) = \frac{D_{ps}}{D_{os}} \sin \hat \alpha
\; , \label{lens1}
\end{equation}
where $D_{ps}$ ($D_{os}$) stands for the angular distance between
the point P close to the lens and the source (the observer and the
source). Also, from the geometry of the lens we have $\sin
\vartheta = b/ D_{ol}$. Hence, choosing $\vartheta$ and the
distance, we have $b$, and $\Delta \varphi$ may be computed
afterwards. In the numerical program, we use again the
dimensionless quantities of Eqs.~(11-12) and instead of the impact
parameter $b$, we follow \cite{DABROWSKI} and take $\vartheta$.
The term $(r/b)^2$ in (\ref{deltaphi}) is then $x^2/(\varpi^2
D_{ol}^2 \sin^2\vartheta )$ or just $x^2/(\varpi^2 D_{ol}^2
\vartheta^2 )$ for small $\vartheta$, respectively. Our numerical
program uses always the correct $\sin\vartheta$ without any
abbreviations so that also angles in the degree regime can be
calculated. The examples in Figs.~\ref{fig9} and \ref{fig10} apply
$\vartheta$ in arc-seconds having the additional unit factor
1/206265 for one arc-second in radians. The change to other units
can then be explained by an additional {\em distance factor} $n$.
For instance, if $\vartheta$ has to be measured in
milli-arc-seconds, $n$ equals $10^{-3}$. In order to get rid of
the distance factor within the numerical program, it is chosen
$D_{ol}=206265/(\varpi n)$.

Furthermore, the reduced angular deflection angle is defined to be
\begin{equation}
\alpha = \vartheta - \beta = \arcsin \left(\frac{D_{ps}}{D_{os}} \sin
\hat \alpha \right) \; .
\end{equation}

A second lens equation can be derived for large deflection angles,
where $D_{ls}$ cannot be considered as being similar to $D_{ps}$.
Of course, by its construction the source is always within a plane
with constant distance to the observer, and studying the
diagramatic view depicted in Fig.~9, it can be obtained that (for
a detailed derivation see Appendix of Ref.\cite{DABROWSKI}),
\begin{eqnarray}
\sin{\alpha} & = & \frac{D_{ls}}{D_{os}} \cos{\vartheta} \cos
\left[ \mbox{arcsin} \left ( \frac{D_{os}}{D_{ls}} \sin (\vartheta
- \alpha ) \right) \right] \times \nonumber \\ & & \times
\left[\tan{\vartheta} + \tan(\hat{\alpha} - \vartheta) \right] \,
,
\end{eqnarray}
where $D_{ls}$ stands for the angular distance between the lens
and the source. Lens equation (\ref{lens1}) requires only the proportion
between $D_{ps}$ and $D_{os}$ so that, in general, the position of source S
describes a more complicated surface.
For the physical situation considered in our
paper, the differences for $\alpha$ amounts a few parts per
thousand of a degree at most so that our Figs.~\ref{fig9} and
\ref{fig10} describe both cases.

Assuming that the boson star lens is half-way between the observer
and the source, such that $D_{ls}/D_{os}=1/2$ and
$D_{ps}/D_{os}=1/2$, we performed numerical computations of the
reduced deflection angle for our new potentials, which we show in
Figs.~\ref{fig9} and \ref{fig10}. The difference among these cases
and the simple mass term (corresponding to the mini-boson star) is
clearly observable. We have taken for the plot the maximum central
density (which produces the maximum deflection angle). In the case
of the mini-boson star, the biggest possible value of $\alpha$ is
23.03 degrees with an image at about $\vartheta = n \times 2.88$
arc-secs with the distance factor $n=n(D_{ol},\Omega)$ which is a
function of the distance from the observer to the lens and the
scalar field frequency, whose inverse can be associated with the
star radius. In our examples we assume that $n=1$, which fixes
$\vartheta$ to be measured in arc-sec.

The characteristics of the boson stars produce the deflection
angles which depend on the observer-to-lens-distance $D_{ol}$ and
the mass. As mentioned above, if $\vartheta$ is chosen to be of
the order of arc-secs (distance factor $n=1$), then the distance
$D_{ol}$ is measured in units of 206265$/\varpi$. Under the
assumption that the mass of the boson star is 10$^{10}$M$_{\odot}$
one has $\varpi \sim 10^{-15}$cm$^{-1}$. Then, the distance
$D_{ol}$ is about 100pc. If the distance factor is $n=10^{-3}$,
and so $\vartheta $ measured in milli-arc-secs, the
boson-star-lens is at about 100kpc.

We assumed that the boson star is transparent and calculated the
deflection angles. All qualitative features of a non-singular
spherically symmetric transparent lens can be revealed using
Figs.~\ref{fig9} and \ref{fig10}: the lens curve for the maximal
boson star. Three images exist, two of them being inside the
Einstein radius and one outside. An Einstein ring with infinite
tangential magnification (tangential critical curve) is found, and
also a radial critical curve for which two internal images merge.
The appearance of the radial critical curve distinguish boson
stars from other extended and non-transparent lenses. For a black
hole or a neutron star, the radial critical curve does not exist
because it is inside the event horizon or the star. Two bright
images near the center of the boson star and the third image at
some very large distance from the center are found. For
non-relativistic cases smaller angles will be found. An
interesting point can be made if one considers an extended source.
In such a case one finds the two radially and tangentially
elongated images very close to each other. Then, looking along the
line defined by these two images the third one can be detected at
a very large distance.

For the Cosh-Gordon potential, we obtain as maximal reduced deflection
angle 23.229 degrees, and for $U_{\exp}$, an even larger deviation, the
biggest possible being 24.391 degrees with an image about at the same place.
Differences among these cases and the usual one is between 0.2 and 0.4
degrees.

\section{Discussion}

In a recent communication, Ho et al.~\cite{HO}
have studied the maximum masses of boson stars
formed with different self-interaction terms
(all of them, however, of power law form and of positive sign).
By just comparing the order of magnitude of the terms involved
in the self-interaction with the mass term, and asking
for them to be of the same order, they were able to see that the
contribution of higher order terms goes as power of
$(m/M_{\rm Pl})^2$. The fact that the contribution to the maximum masses
of these different boson stars decreases (if $m < M_{\rm Pl}$), does
not automatically yield to unobservable effects, as we have discussed in the
previous sections.

The star masses maintain the order of magnitude, for equal single
boson masses, when compared to those cases studied by CSW
\cite{COLPI}, which is in agreement with the results of Ho et
al.~\cite{HO}. This also stems from the fact that all Lagrangians
analyzed are not non-topological ones. Changes are of order of
several percent of the star mass.

For the first time, we have investigated the systems of differential equations
of Einstein-Cosh-Gordon, Einstein-Sine-Gordon, and Einstein-$U(1)$-Liouville.
The different potentials studied so far showed similar gravitational
redshifts and rotational curves, with high angular velocities,
when compared among them,
and between them and the CSW's case. However, large
deviations in the maximum angular deflection have appeared, with differences
amounting appreciable parts of a degree. Some observational effects
distinguish boson stars from other non-transparent compact objects.

However, fair is to say that if an observable determination proves
the existence of a boson star, the effective form of the
Lagrangian may be hidden within the percentage of possible errors.
In that case, facing with the problem of degeneracy
--i.e.~different physical theories giving the same observational
effects-- Occam's razor would probably lead us to consider just
the CSW's choice. Only a detailed knowledge of the boson involved,
and the average form of the interactions, all of them encompassed
in the self-interaction term, may shed light on the explicit model
for the Lagrangian.

The definition of the actual boson which participates in the
construction of the boson star may also influence the transparent
consideration for gravitational lensing phenomena. For instance,
if the star is made up of Higgs particles, we may expect some kind
of interaction apart from the gravitational one that may yield the
star to be a non-transparent object.

Furthermore, we become aware of the possibility that some potentials may give
rise of tunneling of parts of the scalar field. Of course, this effect can
only occur in the quantum regime, hence, if boson stars are in
the order of magnitude of atoms or even atomic nuclei. Our first preliminary
results for the Newtonian case show that especially the form of potentials of
Lee et al.~and of Sine-Gordon can lead to instability due to tunneling.
The effect could mean two things:
(i) The boson (soliton) star is destroyed: it disperses or it forms a black
hole. (ii) The boson (soliton) star experiences an internal rearrangement.
We expect to report on these issues on a forthcoming article.

\section*{Acknowledgments}
We would like to thank Salvatore Capozziello, Mariusz
D\c{a}browski, Gaetano Lambiase, Eckehard Mielke, and Andrew
Whinnett for discussions and comments. D.F.T.~was supported by
CONICET as well as by funds granted by Fundaci\'on Antorchas. He
thanks the hospitality provided by the International Centre of
Theoretical Physics at Trieste and the Universit\'a degli Studi at
Salerno during the latest stages $\vartheta$ of this research.

\begin{figure}[t]
\centering \epsfxsize=8cm \epsfbox{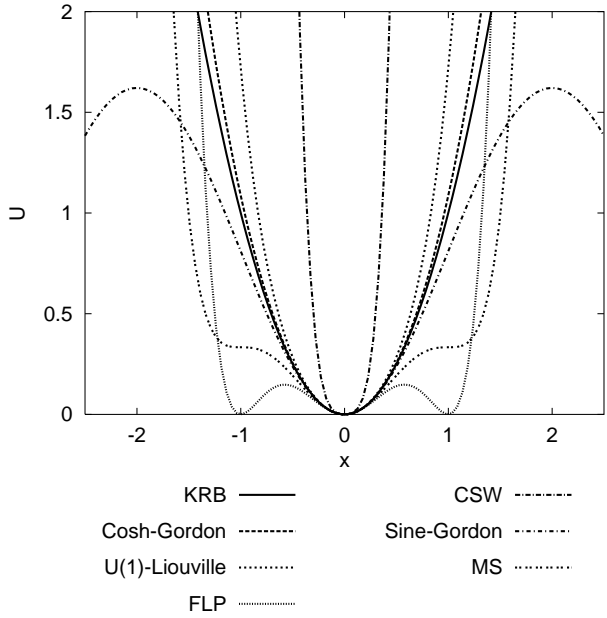}
\caption[fig1]{\label{fig1} Comparison of different potentials.
KRB (Kaup-Ruffini-Bonazzola): $x^2$, Cosh-Gordon: $2(\cosh(x)-1)$,
$U(1)$-Liouville: $\exp (x^2) -1$, FLP (Friedberg-Lee-Pang):
$x^2(1-x^2/2)^2$, CSW (Colpi-Shapiro-Wasserman): $x^2+100 x^4/2$,
Sine-Gordon: $(8/\pi^2) \{ \sin (\pi /2 [x-1])+1 \}$, MS
(Mielke-Scherzer): $x^2-x^4+x^6/3$.   }
\end{figure}

\begin{figure}
\centering \leavevmode \epsfxsize=8cm \epsfbox{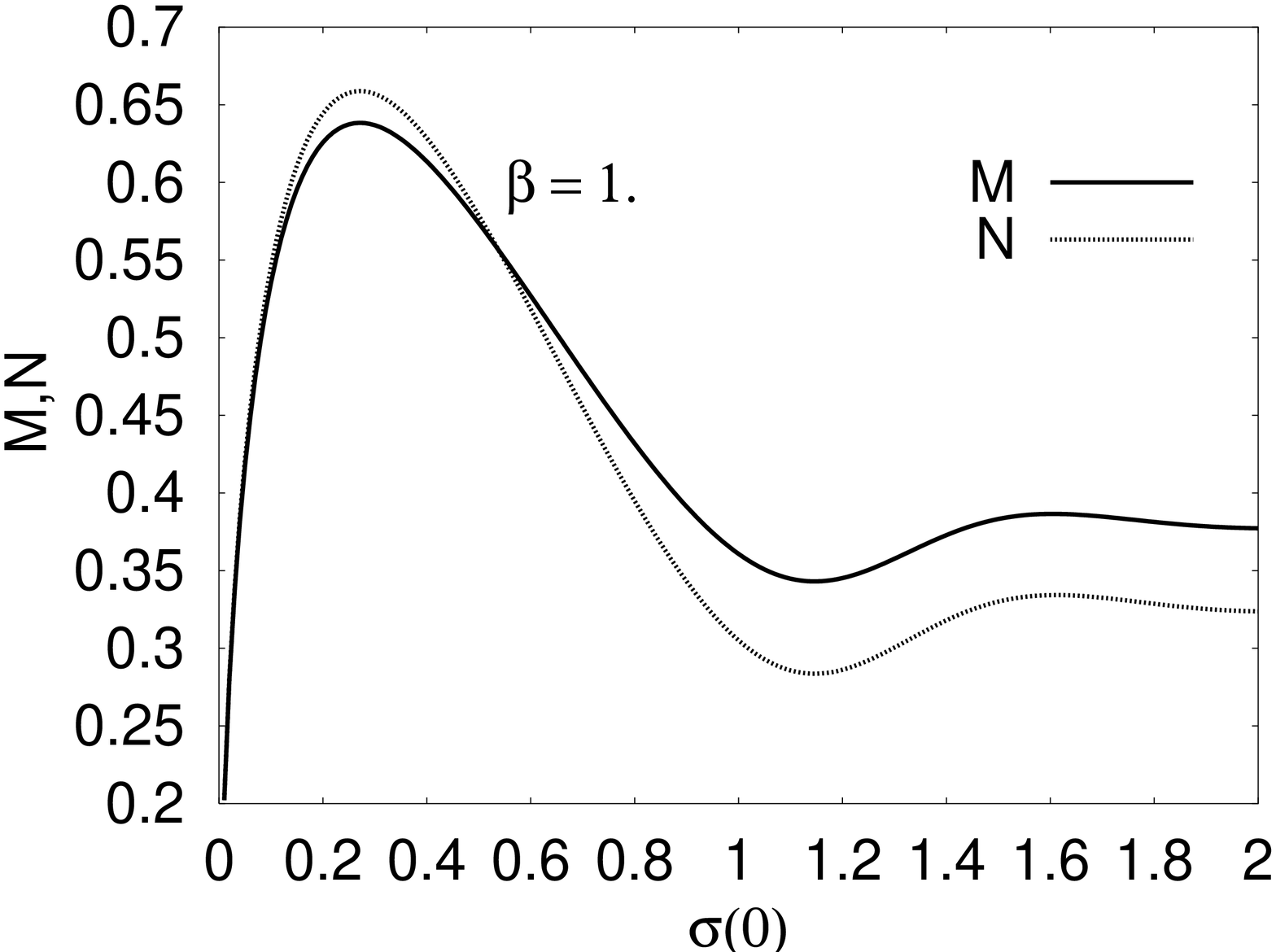}
\caption[fig2]{\label{fig2} Configurations for boson
stars self-interacting via a Cosh-Gordon potential for $\beta=1$.}
\end{figure}

\begin{figure}
\centering \epsfxsize=9cm \epsfbox{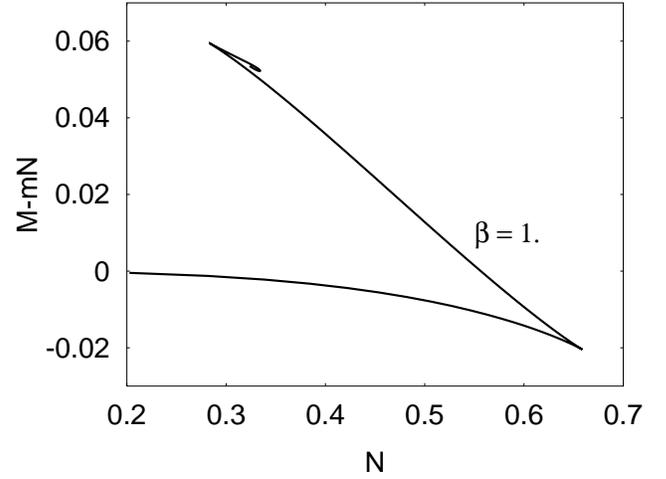}
\caption[fig3]{\label{fig3} Stability analysis for
Cosh-Gordon configurations.}
\end{figure}

\begin{figure}
\centering \leavevmode\epsfxsize=8cm \epsfbox{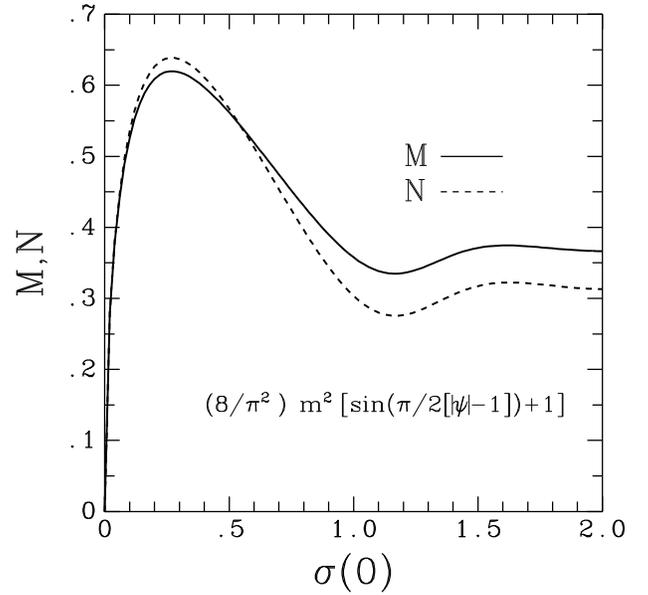}
\caption[fig4]{\label{fig4} Configurations for boson stars
self-interacting via a Sine-Gordon potential. }
\end{figure}

\begin{figure}
\centering \leavevmode\epsfxsize=8cm \epsfbox{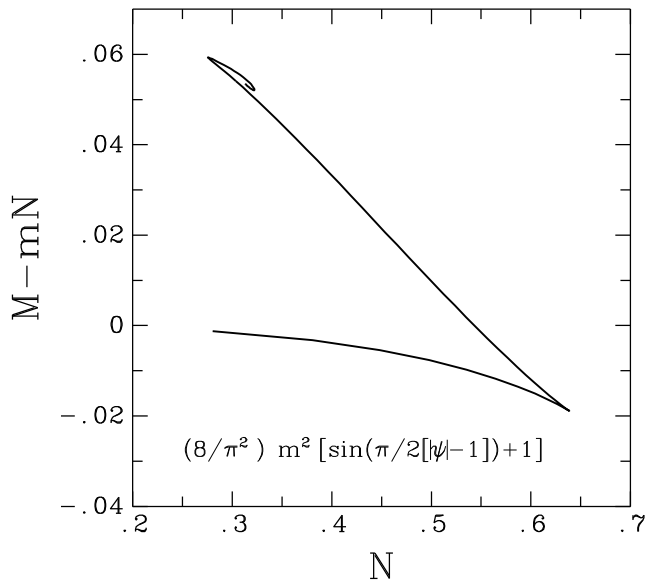} 
\caption[fig5]{\label{fig5} Bifurcation diagram for the
Sine-Gordon potential configurations.}
\end{figure}

\begin{figure}
\centering \leavevmode\epsfxsize=8cm \epsfbox{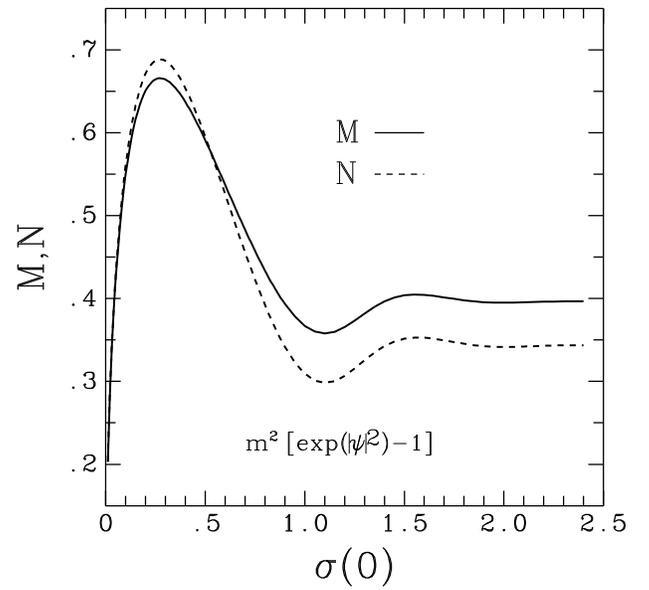} 
\caption[fig6]{\label{fig6} Configurations for boson stars
self-interacting via a $U(1)$-Liouville potential.}
\end{figure}

\begin{figure}
\centering \leavevmode\epsfxsize=8cm \epsfbox{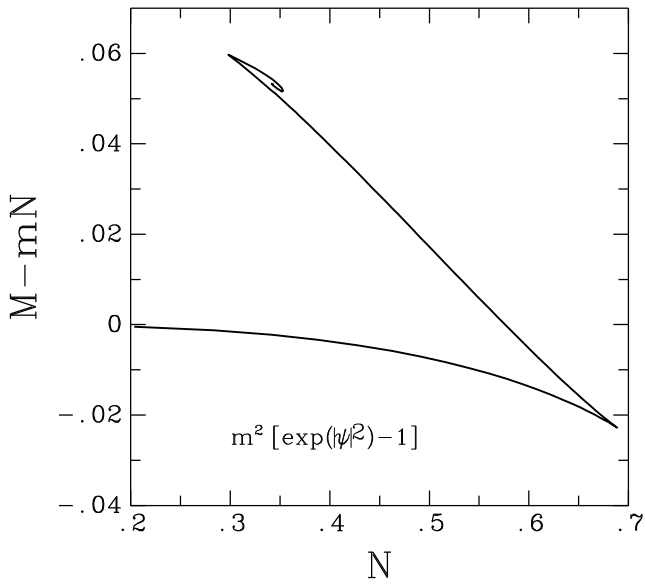} 
\caption[fig7]{\label{fig7} Stability analysis for the
$U(1)$-Liouville potential configurations. The binding energy
changes sign and the presence of cusps signals changes in
stability. The first branch is the only stable one.}
\end{figure}

\begin{figure}
\centering \leavevmode\epsfxsize=8cm \epsfbox{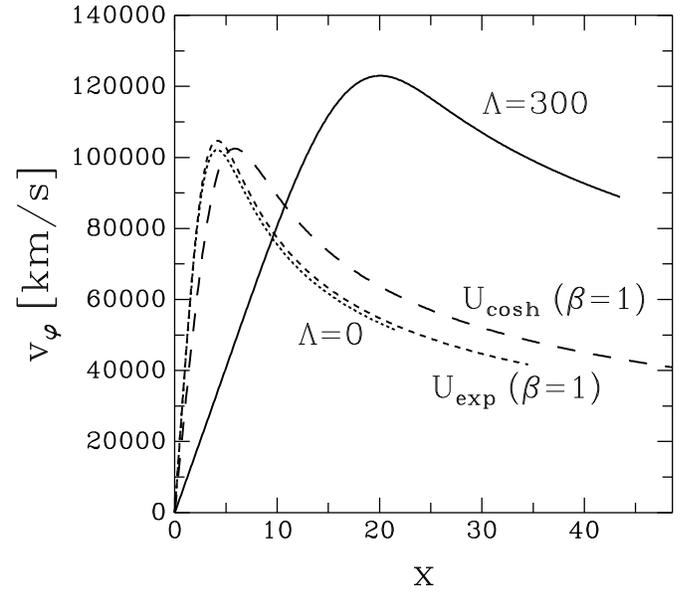} 
\caption[fig8]{\label{fig8} Rotation curves for non-interacting
boson stars and boson stars with generic potentials. Shown in the
Figure are curves corresponding to $\Lambda=0,300$ of the CSW's
choice, taken from the work of Schunck and Liddle, and our new
results for the potentials $U_{{\cosh }}$ and $U_{\exp}$. The
maximal velocities are: 122 990km/s at $x=20.1$ for $\Lambda=300$,
102 073km/s at $x=4.1$ for $\Lambda=0$, 104 685km/s at $x=4.2$ for
$U_{\exp}$, and 102 459km/s at $x=5.9$ for $U_{{\cosh }}$. }
\end{figure}

\begin{figure}
\centering \leavevmode\epsfysize=12cm \epsfbox{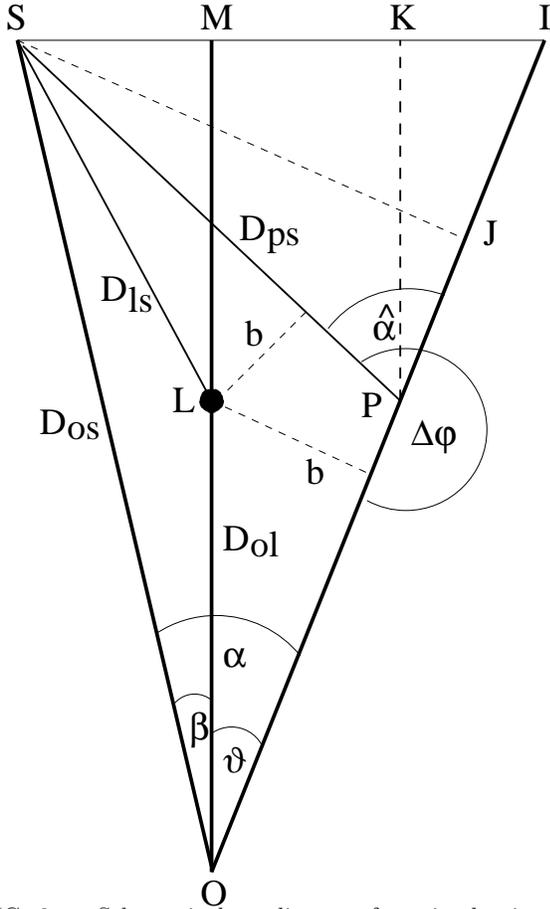} 
\caption[fig9a] {\label{fig9a} Schematic lens diagram for
microlensing phenomenon. $D_{ol}$ is the distance from the
observer (O) to the lens (L), $D_{os}$ is the observer-source
distance, and $D_{ls}$ the distance from the lens to the source
(S). The angle $\beta$ denotes the true angular position of the
source whereas $\vartheta $ measures the angle of the image
position. $D_{ps}$ is the distance from the point P in the source
plane to the source (for small deflection angles, $D_{ps} \sim
D_{ls}$, as well as $D_{os} \sim D_{ol} + D_{ls}$). For the
meaning and use of all other parameters see text and the Appendix
of the paper by D\c{a}browski and Schunck.}
\end{figure}

\begin{figure}
\centering \leavevmode\epsfxsize=8cm \epsfbox{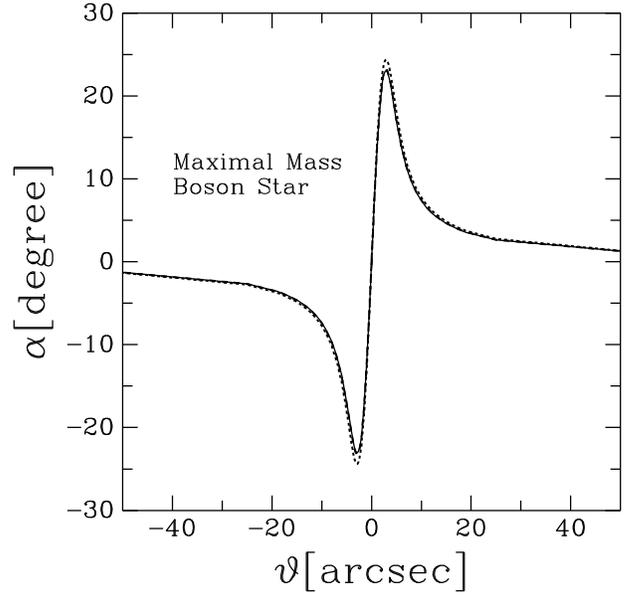} 
\caption[fig9]{\label{fig9} The reduced deflection angle
(difference between the true and the image angular position) as a
function of the the image position for the different potentials.
$U_{\exp}$ produces the largest deflection angle in comparison
with the simple mass term potential (CSW's choice with
$\Lambda=0$) and the Cosh-Gordon potential. We have chosen here
the stable maximal mass boson star.}
\end{figure}

\begin{figure}
\centering \leavevmode\epsfxsize=8cm \epsfbox{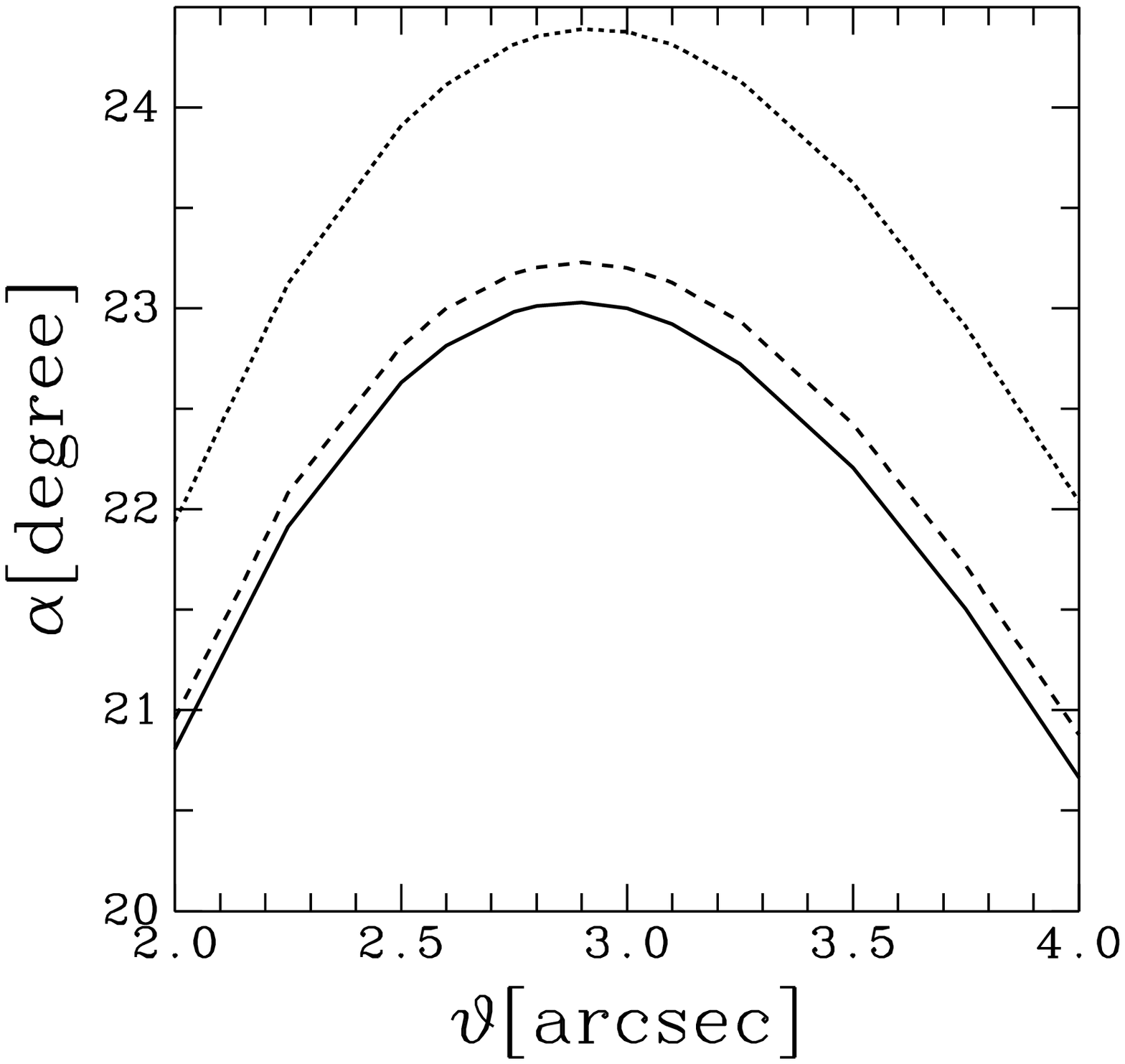} 
\caption[fig10]{\label{fig10} Comparison for all three potentials
at the maximal reduced deflection angle. The upper curve is for
$U_{\exp}$, the middle curve is for $U_{{\cosh }}$ and the lower
one for $\Lambda=0$ CSW's choice.}
\end{figure}


\begin{references}
\bibitem{KAUP} D. J. Kaup, Phys. Rev. {\bf 172}, 1331 (1968).

\bibitem{RB} R. Ruffini and S. Bonazzola, Phys. Rev. {\bf 187}, 1767 (1969).

\bibitem{MS81} E.W. Mielke and  R. Scherzer, Phys. Rev. D{\bf 24}, 2111 (1981).

\bibitem{COLPI} M. Colpi, S. L. Shapiro and I. Wasserman,
Phys. Rev. Lett. {\bf 57}, 2485 (1986).

\bibitem{SCHUNCK} F. E. Schunck, ``A scalar field matter model for dark halos
of galaxies and gravitational redshift'', astro-ph/9802258.

\bibitem{DABROWSKI} M. P.~D\c{a}browski and F. E.~Schunck,
``Gravitational lensing of boson stars'', astro-ph/9807207,
accepted by Astrophys. J.

\bibitem{RBS} F. E.~Schunck and E. W.~Mielke, Phys.~Lett.
A{\bf 249}, 389 (1998).

\bibitem{reviews} P. Jetzer, Phys. Rep. {\bf 220}, 163 (1992); A. R.
Liddle and M. S. Madsen, Int. J. Mod. Phys. D{\bf 1}, 101 (1992);
E.W.~Mielke and F.E.~Schunck: ``Boson stars: Early history and
recent prospects'',
Proceedings of the 8th Marcel Grossmann meeting in Jerusalem,
(World Scientific Publ., Singapore 1999), gr-qc/9801063.

\bibitem{TORRES-BOSON} D. F. Torres, Phys. Rev. D{\bf 56}, 3478 (1997).

\bibitem{LIDDLE-SCHUNCK} F. E. Schunck and A. R. Liddle, Phys. Lett.
B{\bf 404}, 25 (1997).

\bibitem{MEMORY} D. F.~Torres, A. R.~Liddle, and F. E.~Schunck,
Phys. Rev. D{\bf 57}, 4821 (1998);
D. F.~Torres, F. E.~Schunck, and A. R.~Liddle,
Class.~Quantum Grav. {\bf 15}, 3701 (1998).

\bibitem{WHINNETT}
A. W. Whinnett and D. F. Torres, Phys. Rev. D{\bf 60}, 104050
(1999).

\bibitem{L87} T.D. Lee,  Phys. Rev. D{\bf 35}, 3637 (1987).

\bibitem{FLP87}
R. Friedberg, T.D. Lee, and Y. Pang, Phys. Rev. D{\bf 35} 3640, 3658,
3678 (1987).

\bibitem{lee}
T.D.~Lee and Y.~Pang,  Phys.~Rep. {\bf 221}, 251 (1992).

\bibitem{BOOK} T. D. Lee, {\it Particle physics and introduction
to field theory}, Chapter 7,
(Harwood Academic Publishers, Chur, Switzerland 1981).

\bibitem{KUSMARTSEV} F. V.~Kusmartsev, E. W.~Mielke, and F. E.~Schunck,
Phys.~Rev. D{\bf 43}, 3895 (1991).

\bibitem{KUSMARTSEV2} F. V.~Kusmartsev, E. W.~Mielke, and F. E.~Schunck,
Phys.~Lett. A{\bf 157}, 465 (1991).

\bibitem{KUSMARTSEV3} F. V.~Kusmartsev and F. E.~Schunck,
Physica B{\bf 178}, 24 (1992).

\bibitem{TOLMAN}
R.C. Tolman, Phys. Rev. {\bf 35} (1930) 875;
R.C. Tolman, {\it  Relativity, Thermodynamics and Cosmology}
(Clarendon, Oxford, 1934).

\bibitem{VIBRA} K. S. Virbhadra, D. Narashima, and S. M. Chitre,
A \& A {\bf 337}, 1 (1998).

\bibitem{TORR}
J. G. Cramer, R. L. Forward, M. S. Morris, M. Visser, G.
Benford, and G. A. Landis, Phys. Rev D{\bf 51}, 3117 (1995),
D. F. Torres, G. E. Romero, and L. A. Anchordoqui,
Phys. Rev. D{\bf 58}, 123001 (1998); {\it ibid.}
Mod. Phys. Lett. A{\bf 13}, 1575 (1998).

\bibitem{SEF92}
P. Schneider, J. Ehlers, and E.E. Falco,
``Gravitational Lenses'' (Springer, Berlin 1992).

\bibitem{MICRO} R. Narayan and M. Bartelmann: ``Lectures on gravitational
lensing", in: Formation of Structure in the Universe, Proceedings of
the 1995 Jerusalem Winter School; edited by A.Dekel and J.P. Ostriker
(Cambridge University Press, Cambridge); astro-ph/9606001.



\bibitem{HO}J. Ho, S. Kim, and B.--H. Lee: ``Maximum mass of boson
star formed by self-interacting scalar fields", gr-qc/9902040.


\end{references}
\end{document}